\documentclass[twocolumn,showpacs,preprintnumbers,amsmath,amssymb]{revtex4}

\usepackage{epsfig}
\usepackage{graphicx}
\usepackage{dcolumn}
\usepackage{bm}
\def\pt#1{\phantom{#1}}
\newcommand{\oo}{{\omega_\oplus}}
\newcommand{\om}{\omega}
\newcommand{\Om}{\Omega}

\newcommand{\Oo}{\Omega_\oplus}

\newcommand{\be}{\begin{equation}}
\newcommand{\ee}{\end{equation}}

\newcommand{\cov}{{\rm cov}}
\newcommand{\is}{i_4\bar s}
\newcommand{\kemin}{\tilde\kappa_{e-}}
\newcommand{\kopl}{\tilde\kappa_{o+}}

\begin{document}

\preprint{APS/123-QED}

\title{Atom interferometry tests of local Lorentz invariance in gravity and electrodynamics}
\author{Keng-Yeow Chung}
\affiliation{Physics Dept., National University of Singapore, 2
Science Drive 3, Singapore 117542}
\author{Sheng-wey Chiow}
\affiliation{Physics Department, 382 Via Pueblo Mall, Stanford, CA
94305, USA}
\author{Sven Herrmann}
\affiliation{ZARM - University of Bremen
Am Fallturm/Hochschulring, 28359 Bremen,
Germany}
\author{Steven Chu}
\affiliation{Department of Physics, University of California,
Berkeley, CA 94720}
\affiliation{Lawrence Berkeley National
Laboratory, One Cyclotron Road, Berkeley, CA 94720}
\author{Holger M\"uller}
\email{hm@berkeley.edu} \affiliation{Department of Physics,
University of California, Berkeley, CA 94720}

\date{\today}

\begin{abstract}
We present atom-interferometer tests of the local Lorentz
invariance of post-Newtonian gravity. An experiment probing for
anomalous vertical gravity on Earth, which has already been
performed by us, uses the highest-resolution atomic gravimeter so
far. The influence of Lorentz violation in electrodynamics is also
taken into account, resulting in combined bounds on Lorentz
violation in gravity and electrodynamics. Expressed within the
standard model extension or Nordtvedt's anisotropic universe
model, we limit twelve linear combinations of seven coefficients
for Lorentz violation at the part per billion level, from which we
derive limits on six coefficients (and seven when taking into
account additional data from lunar laser ranging). We also discuss
the use of horizontal interferometers, including atom-chip or
guided-atom devices, which potentially allow the use of longer
coherence times in order to achieve higher sensitivity.
\end{abstract}

\pacs{04.80.Cc, 03.75.Dg, 04.25.Nx, 11.30.Cp}
\maketitle


\section{Introduction}
Local Lorentz invariance (LLI) in the gravitational interaction
can be viewed as a prediction of the theory of general relativity. And it is not a trivial consequence, given
that alternative theories of gravity have been put forward that do
not lead to LLI, yet agree with general relativity in their
predictions for the red-shift, perihelion shift, and time delay.
Experimental tests of the LLI in gravity are required to decide
between these theories \cite{Will71}.

Another reason to perform tests of LLI in gravity is connected to
one of the outstanding problems in physics, to find a unified
theory of quantum gravity. The natural energy scale for such a
theory is the Planck scale of about $10^{19}\,$GeV. Direct
experimentation at the Planck scale is, unfortunately, not
possible. However, it is possible to search for suppressed effects
at attainable energy scales in experiments of outstanding
sensitivity. Violations of LLI (``Lorentz violation") are among
the relatively few candidates for such observable consequences of
quantum gravity \cite{AmelinoCamelia}.

The LLI of the non-gravitational standard model has been tested
for various particles, including photons
\cite{KosteleckyMewes,KosteleckyMewesPRD,photontests,Mueller07},
electrons
\cite{electrontests,Kornack,Hunter,KosteleckyLane,Mueller07,Mueller03,Lane},
protons \cite{protontests,Kornack,KosteleckyLane}, neutrons
\cite{Kornack,Hunter,KosteleckyLane,neutrontests}, and others (see
\cite{AmelinoCamelia,Mattingly,WillReview,WillBook,Tables} for
recent reviews). However, the LLI of gravity itself has been
studied little to this date. Thus, it remains interesting and
important to explore the validity of LLI in gravity.

For the theoretical descriptions of the consequences of Lorentz
violation in gravity in the solar system, a post-Newtonian
approximation is justified. Such descriptions are known as
Nordtvedt's anisotropic universe model \cite{Nordtvedt76} or the
standard model extension (SME)
\cite{ColladayKostelecky,Kostelecky,BaileyKostelecky}, see Sec.
\ref{hypsignal} for details. These frameworks use the following
Lagrangian to describe the interaction between two point-masses
$M$ and $m$ \cite{BaileyKostelecky}
\begin{eqnarray}\label{Lagrangian}
\mathcal L=\frac12  m v^2+
G\frac{Mm}{2r}\left(2+3 \bar s^{00}\right. \nonumber \\
\left.+ \bar s^{jk} \hat r^j\hat r^k -3\bar s^{0j} v^j-\bar
s^{0j}\hat r^j v^k \hat r^k\right).
\end{eqnarray}
For simplicity, we have taken $M$ to be at rest. We denote $\vec
r$ the separation between $M$ and $m$, pointing towards $m$. The
indices $j,k$ denote the spatial coordinates, $\vec v$ the
relative velocity, and $\hat r=\vec r/r$. The components of $\bar
s^{\mu\nu}=\bar s^{\nu\mu}$ specify Lorentz violation in gravity.
If they vanish, LLI is valid.

The relative weakness of gravity means that only a small set of
exceptionally sensitive experiments can place interesting limits
on $\bar s$. Nordtvedt and Will noted that Lorentz violation in
gravity would cause a modulation of the apparent local
gravitational acceleration $g$ as the Earth rotates in space.
Using gravimeter data \cite{Harrison63} taken during the
international geophysical year, July 1957-December 1958, they
obtained a limit of $|\bar s^{JK}|\leq 4\times 10^{-9}$ ($J,K\in
\{X,Y,Z\}$) on the spatial components
\cite{Will71,Nordtvedt76,NordtvedtWill72}. More recently, Battat
{\em et al.} analyzed 34 years of lunar laser ranging data,
finding bounds on two linear combinations of the $\bar s^{JK}$ at
a level of $\sim 10^{-10}$ \cite{Battat}. They also found bounds
on the three $\bar s^{TJ}$ at levels of $\sim 10^{-7}$.

This means that for four degrees of freedom of $\bar s^{JK}$,
there has been no improvement for about five decades. Moreover, a
fundamental issue remains unaddressed: testing the isotropy of any
force of nature means comparing it to another `standard' that is
assumed to be isotropic. In other words, the isotropy of gravity
can only be tested against the isotropy of another phenomenon,
such as the velocity of light $c$. Past tests, however, are not
easy to analyze in such depth as to make this comparison explicit.
For example, it would be very difficult to analyze an influence of
the isotropy of $c$ on the spring gravimeters used by Nordvedt and
Will, although a description of Lorentz violation in solids is
possible in principle \cite{ResSME,H2SME,TLVMC}. Hence, it is
desirable to perform an experiment that relies on sufficiently
simple physical principles so that all channels of influence for
Lorentz violation can be theoretically described.

\begin{figure}
\centering \epsfig{file=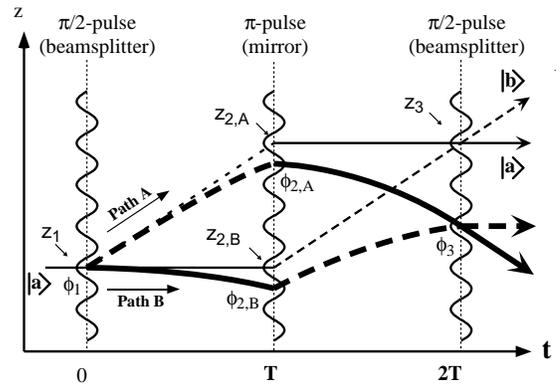,width=0.4\textwidth}
\caption{Space-time diagram of a Mach-Zehnder atom interferometer.
$t$ denotes time, $z$ the vertical coordinate. \label{MZschem}}
\end{figure}

With atom interferometry
\cite{KasevichChu91,Chureview,Pritchardreview}, we have the unique
situation of an experiment which is not only sensitive enough to
obtain improved bounds on several elements of $\bar s^{JK}$, but
also `clean' in the sense that we can quantitatively understand
the influence of Lorentz violation in all relevant sectors. The
basic principle is shown in Fig. \ref{MZschem}: A matter wave
packet is split by a beam splitter to form two interferometer
arms. Because of the interaction with external potentials, such as
gravity, the wave packet picks up a phase difference $\varphi$
between the arms. When the arms are recombined at a final beam
splitter, $\varphi$ determines the probability that the atom is
found emerging from one of the two interferometer outputs.
Gravity's contribution to $\varphi$ can be several $10^7$ radians
for the most sensitive devices \cite{LeGouet,LVGrav}.

Light-pulse atom interferometers use standing waves of laser light
as gratings to diffract the matter waves; since the period of
these gratings is given by the laser wavelength, the extremely
high accuracy of laser frequency and phase stabilization can be
applied to the measurement of gravity and other inertial forces.

Atom interferometers have been used for measurements of the
fine-structure constant $\alpha$
\cite{Weiss,Wicht,Biraben,Paris,BraggInterferometry}, the local
gravitational acceleration $g$ \cite{Peters}, its gradient
\cite{Snaden98}, the Sagnac effect \cite{Gustavson,Durfee,Canuel},
or Newton's gravitational constant \cite{Fixler,Lamporesi}. They
rival or exceed the performance of other state-of-the-art methods.
Thus, it seems natural to apply atom interferometry as a device to
probe the weakest of the known forces of nature
\cite{KasevichChu91}. Detailed proposals have already been put
forward, including tests of the Einstein Equivalence principle
\cite{Dimopoulos} and detection of gravitational waves
\cite{GravWav}.

Here, we will discuss the first atom-interferometer test of
post-Newtonian gravity \cite{LVGrav}. It is also one of the few
laboratory tests of gravity whose sensitivity is competitive with
the best astrophysics data. This paper is organized as follows: We
start with a description of the experiment in Sec. \ref{exp};
details not covered there will be found elsewhere
\cite{LVGrav,Peters,Hensleyvib,Treutlein,Kengyeow}. Sections
\ref{prin} through \ref{inter} describe the setup; section
\ref{hypsignal} describes the hypothetical signal for Lorentz
violations in this experiment, including the influence of the
electromagnetic sector. Section \ref{deconvolution} gives details
of the data analysis. In Sect. \ref{Hinterf}, we consider
possibilities for future tests based on horizontal interferometers
that may be based on atom chips \cite{Hoffenberth,Wang,Shin} or
other matter waveguides\cite{Wu,Gupta05}.

\section{Experiment: a vertical atomic fountain interferometer} \label{exp}
\subsection{Principle}\label{prin}

In our Mach-Zehnder atom interferometer (Fig. \ref{MZschem}), two
vertical and antiparallel laser beams make the beam splitters and
mirrors of the interferometer. They drive two photon Raman
transitions: the atom absorbs a photon from one beam and is
stimulated to emit a photon into the other one. Thus, the atom
changes its hyperfine state and receives the momentum $2\hbar k$
of two photons. A `$\pi/2$' pulse, which has its intensity and
duration chosen such that this process happens with a probability
of 1/2, acts as a beam splitter; a `$\pi$' pulse, with a
transition probability close to one, makes a mirror.

For our Mach-Zehnder interferometer, a first $\pi/2$ laser pulse
transfers the atoms into a superposition of the $F=3, m_F=0$ and
$F'=4, m'_F=0$ states. These move vertically relative to each
other because of the momentum transferred by the laser radiation.
A total of three light pulses split, reflect, and then recombine
the paths to form an interferometer. They are separated in time by
the pulse separation time $T_p$.

The matter waves in both paths acquire a relative phase difference
$\varphi=S_{Cl}/\hbar+\varphi_I$ (see, e.g., Ref. \cite{Peters}
for details). The phase of the free evolution of the wave packet
between the beam splitters is given by the classical action
$S_{Cl}$. If we restrict our attention to a constant gravitational
acceleration $g$,
\begin{equation} \frac
{S_{Cl}}\hbar=\frac 1 \hbar \int_0^{2T_p}\left[\frac 12 m \dot
z^2-mgz\right]\,dt\,.
\end{equation}
$z$ is the vertical coordinate and $m$ the atom's mass. By
calculating the integral over the classical trajectories, it can
be shown that this is the same for the upper and lower trajectory.
Thus, it does not contribute to the phase difference between the
interferometer arms.

The phase $\varphi_I$ is because whenever the atom changes state
during an interaction, the phase of the atom changes by an amount
equal to the phase of the light field. This adds phase when the
atom absorbs a photon and subtracts phase when the atom emits one.
Thus, it is easy to see that $\varphi_I=0$ for the trajectories
without gravity, depicted by light lines in Fig. \ref{MZschem}.
Gravity accelerates the atoms downwards according to $\Delta
z=-gt^2/2$, as depicted by heavy lines.  This gives rise to a
phase shift of $k_{\rm eff}\Delta z$, where $k_{\rm eff}$ denotes
the effective wavenumber. To high accuracy, the laser beams can be
modeled as plane waves, which for two-photon transitions results
in an effective wavenumber of $k_{\rm eff}=k_1+k_2$, where
$k_{1,2}$ are the wavenumbers of both lasers. For the upper
trajectory, a photon pair is absorbed at $t=0$ and emitted at
$t=T_p$. We can set the phase of that interaction to zero by
definition. The phase of the second one will then be $k_{\rm
eff}gT_p^2/2$. For the lower path, a pair is absorbed at $t=T_p$
and one emitted at $2T_p$. They add to a phase of $-k_{\rm
eff}gT_p^2/2+k_{\rm eff}g(2T_p)^2/2=3k_{\rm eff}gT_p^2/2$. The
difference for the two paths is thus $\varphi_I=k_{\rm
eff}gT_p^2$. For the total phase shift, we take into account the
possibility that the phases of the laser pulses at $z=0$ can
experimentally be set arbitrary values
$\varphi_1,\varphi_2,\varphi_3$ by suitable control of the laser
system. We obtain \cite{Peters}
\begin{eqnarray}\label{phase}
\varphi&=&k_{\rm eff}gT_p^2-\varphi_L,\nonumber\\
\varphi_L&=&\varphi_1-2\varphi_2+\varphi_3.
\end{eqnarray}
For our experiment, $T_p=0.4\,$s and $k\simeq 2\pi$/(852\,nm), so
$\varphi \simeq 2.3 \times 10^{7}$\,rad. In the experiment, we set
$\varphi_L=rT_p^2$ by ramping the difference frequencies of the
laser (actually, the continuous ramp can be approximated by
discrete steps, as the laser pulse duration is very short). To
measure $g$, the interferometer phase $\varphi$ is zeroed by
adjusting the ramp rate to $r_0 \approx 2\pi\times 23\,$MHz/s.
This corresponds to finding the center of the interference
pattern. Then, $g=r_0/(k_{\rm eff})$. If $\varphi$ can be measured
to, e.g., 1\,mrad, we obtain a resolution of $10^{-10}$ in $g$.

\subsection{Fountain}

\begin{figure}
\centering \epsfig{file=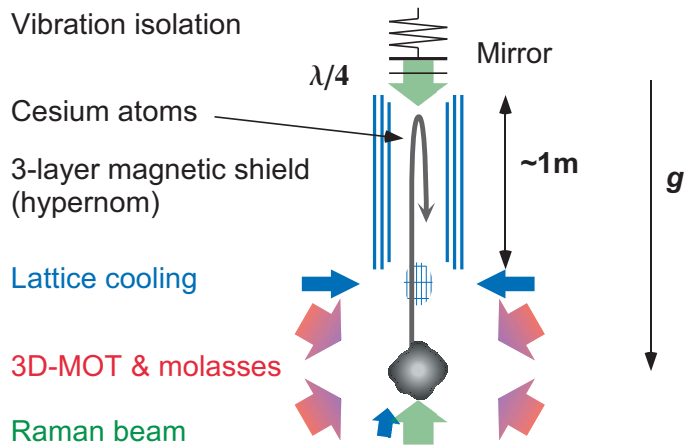, width=0.5\textwidth}
\caption{\label{setup} Setup. $\lambda/4$; $1/4$-wave retardation
plate.}
\end{figure}

In our experimental setup (Fig. \ref{setup}), we assemble about
$10^9$ Cs atoms within 650\,ms from a background vapor pressure of
$\sim 10^{-9}$\,mbar in a 3-dimensional magneto-optical trap
(3D-MOT). A moving optical molasses launch accelerates them
vertically upwards to a $\sim 1-$s ballistic trajectory with a
temperature of 1.2-2$\,\mu$K. Raman sideband cooling in a
co-moving optical lattice results in $\sim 3\times 10^8$ atoms in
the $F=3, m_F=3$ state at a (3D) temperature of 150\,nK
that form a cloud of roughly 3\,mm$^2$ area \cite{Treutlein}. 
A sudden change in the magnetic field followed by a 120-$\mu$s
microwave pulse transfers $\sim 20\%$ of them into the $F=4,
m_F=0$ state. Atoms left over in the $F=3$ state are then cleared
away using a resonant laser pulse. A solenoid generates a small
magnetic bias field to set the quantization axis.

\subsection{Interferometer}\label{inter}

The off-resonant Raman pulses for the beam splitters have a
wavelength of 852\,nm and are generated by two grating-stabilized
extended cavity diode lasers that are based on 100-mW laser diodes
SDL-5411. The first is frequency stabilized (`locked') to a cesium
vapor cell using Doppler-free saturation spectroscopy. It arrives
at the experiment with a detuning of -1030MHz from the $6S_{1/2},
F=3\rightarrow 6P_{3/2}, F'=4$ transition in Cs. The second one is
phase locked to the first one with a frequency difference close to
the hyperfine splitting of $\simeq 9192$\,MHz, referenced to a
LORAN-C frequency standard. The two lasers are overlapped on a
beam splitter and the combined beam, containing 20\,mW of each
one, is transmitted to the experiment via a single-mode,
polarization maintaining optical fiber. There, the beams are
switched and intensity-controlled by an acousto-optical modulator
(Isomet 1205). They are collimated with a $1/e^2$ intensity
diameter of about 2.5\,cm, and pass the vacuum chamber with linear
polarization. Retro-reflection on top of it with two passes
through a quarter-wave retardation plate forms a lin$\perp$lin
polarized counterpropagation geometry (Fig. \ref{setup}).

The error signal for the phase lock is generated by detecting the
beat note after the fiber. Phase noise sources before this point
can thus be taken out by the phase lock loop, whereas most others
are common to both beams and do not affect the interferometer. The
most important exception are residual vibrations of the top
mirror, which we therefore reduce to below $5\times
10^{-9}\,g/\sqrt{\rm Hz}$ in a frequency range of 0.1-10\,Hz by a
sophisticated active vibration isolator \cite{Hensleyvib}.


For fluorescence detection with a Hamamatsu R943-02
photomultiplier tube (PMT), the $F=4$ interferometer output is
driven on a cycling transition; the $F=3$ output is detected
subsequently after optical pumping to the $F=4$ state.
Normalization of the signals takes out variations in the number of
launched atoms.

\begin{figure}
\centering \epsfig{file=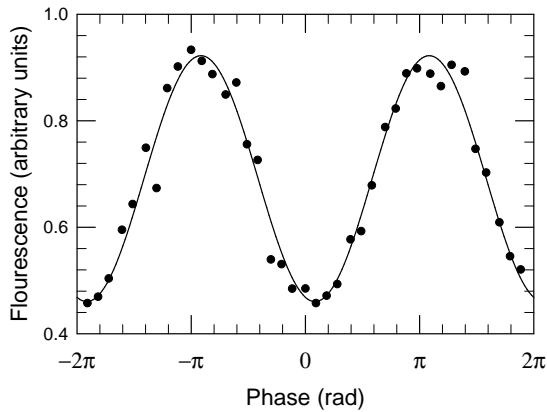,width=0.4\textwidth}
\caption{Typical fringe obtained in our experiment
\label{MZfringe}}
\end{figure}

Fig. \ref{MZfringe} shows a typical gravity fringe with a pulse
separation time of $T_p=400\,$ms, taken with 40 launches that take
75\,s total. The sinewave-fit has a phase uncertainty of
0.031\,rad, and determines $g$ to an uncertainty of $\sim
1.3\times 10^{-9}$\,g. This corresponds to $11\times
10^{-9}\,g/\sqrt{\rm Hz}$. An improved short-term resolution of
$8\times 10^{-9}\,g/\sqrt{\rm Hz}$ can be reached by taking data
at the 50\% points of the fringes only. However, as this method is
more sensitive to systematic effects such as drift of the PMT
sensitivity \cite{Peters}, we used fringe-fitting for taking the
long term data. Our resolution is about three times better than
the best previous one, which was reported by Peters {\em et al.}
\cite{Peters}. This is mainly a consequence of the increased
interaction time $T=400\,$ms. It also surpasses the best classical
absolute gravimeter, the FG-5 falling corner cube gravimeter, by a
factor of about 20.\\

\subsection{Hypothetical Signal}
\label{hypsignal}

The notion that the gravitational force between two objects might
depend on the direction of the separation $\vec r$ could be
described in very simple terms. For this work, however, we want to
use a model that is as general as possible on the one hand and
compatible with accepted principles that underlie the standard
model and gravitational theory on the other hand. Two such models
suggest themselves, Nordtvedt's anisotropic universe model
\cite{Nordtvedt76} and the standard model extension (SME)
\cite{ColladayKostelecky,Kostelecky,BaileyKostelecky} (a simplified formalism with one parameter has also been proposed \cite{Li}.) The SME starts from a Lagrangian formulation of the standard model and
gravity, adding general Lorentz violating terms that can be formed
from the fields and tensors. It is therefore a comprehensive
theoretical framework that allows us to model violations of LLI in
the various sectors of the standard model and gravity.

As $\varphi=k_{\rm eff} gT^2_p$, the hypothetical signal for
violations of LLI can be due to a change in $g$ and $k_{\rm eff}$.
The signal may thus be due to Lorentz violation in both the
electromagnetic as well as the gravitational sector.

\subsubsection{Gravitational sector}

In a post-Newtonian approximation, the Lagrangian for the
gravitational interaction in the SME is given by Eq.
\ref{Lagrangian}. The two-body Lagrangian of the anisotropic
universe model is similar, but $\bar s^{00}=0$ and the
coefficients of $v^j_a$ and $\hat r^j_{ab} v^k_{a} \hat r^k_{ab}$
are independent of each other.

In principle, the components of $\bar s$ can be defined in any
inertial frame of reference. For experiments on Earth (as well as
on satellites), it is convenient to choose a Sun-centered
celestial equatorial reference frame \cite{KosteleckyMewesPRD}. It
has the $X$ axis pointing towards the vernal equinox (spring
point) at 0\,h right ascension and $0^{\circ}$ declination, the
$Z$ axis pointing towards the celestial north pole ($90^{\circ}$
declination) and the $Y$ axis in the way needed to complete the
right handed orthogonal dreibein. Earth's equatorial plane is the
$X-Y$ plane and the orbital plane of the Earth is tilted at an
angle $\eta \simeq 23^\circ$ with respect to the latter. The time
scale $T$ is set by $T = 0$ when the Sun passes the spring point,
which, for example, happened in 2001 on March 20, 13 h 31 min
universal time (UT). Sun--centered frame quantities have Greek
(0-3) or capital Latin indices.

We also define a laboratory frame, which has the $x$ axis pointing
south, the $y$ axis east, and the $z$ axis vertically upwards. The
laboratory time scale is set by $T_\oplus=0$ at any one instant
when the $y$ and the $Y$ axis coincide. The difference between the
two time scales $T_\oplus$ and $T$ can be written as a phase
difference $\phi=\om_\oplus(T_\oplus-T)$ \cite{BaileyKostelecky},
where $\oo$ is the sidereal angular frequency of Earth's rotation;
$\phi\simeq -1.77$ for this experiment. Laboratory--frame
quantities have small Latin indices.

The derivation of the time-dependent modulations of $g$ for an
observer on Earth involves taking into account the rotation and
orbit of the Earth; the Earth itself is modelled as a massive
sphere having a spherical moment of inertia of $I_\oplus\approx
M_\oplus R_\oplus^2/2$ \cite{WillBook} (not to be confused with
the conventional moment of inertia, which for Earth is about
$M_\oplus r_\oplus^2/3$). It suffices to consider the first order
in the Earth's orbital velocity $V_\oplus \simeq 10^{-4}$. Bailey
and Kostelecky \cite{BaileyKostelecky} have studied this in
detail, and we refer the reader to this reference for the detailed
signal components in the purely gravitational sector.

\subsubsection{Electromagnetic sector}
\label{emsector}

To study the variations of $k_{\rm eff}$ caused by Lorentz
violation, we start from the Lagrangian for the electromagnetic
sector of the SME,
\begin{eqnarray}\label{SMEphotonlagrangian}
{\mathcal L} & = & - \frac 14 F^{\mu \nu} F_{\mu \nu} - \frac 14
(k_F)_{\kappa \lambda \mu \nu} F^{\kappa \lambda} F^{\mu \nu},
\end{eqnarray}
where $F^{\mu \nu}$ is the electromagnetic field tensor and
$A^\mu$ the vector potential. The second term is proportional to a
dimensionless tensor $(k_F)_{\kappa \lambda \mu \nu}$, which
vanishes, if Lorentz invariance holds on electrodynamics. The
tensor has 19 independent components \cite{KosteleckyMewesPRD}. A
brief summary of the Maxwell equations that are derived from this
Lagrangian is given in appendix \ref{EMappendix}.

For studying the Lorentz-violating modification to the effective
wavevector $k_{\rm eff}$ in the atom interferometer, the plane
wave solutions have to be found. Making the standard ansatz
$F_{\mu\nu}(x)=F_{\mu\nu}(p)e^{-ik_\alpha x^\alpha}$ for a plane
wave with a wave 4-vector $k^\alpha=(k^0,\vec k)$ and inserting
into Eq. (\ref{inhomMaxwell}) one obtains the dispersion relation,
that will allow us to determine $k_{\rm eff}$ as a function of the
direction of propagation. Let
\begin{eqnarray}
\rho&=&-\frac12 \tilde k_\alpha{}^\alpha, \quad \sigma^2=\frac 12
(\tilde k_{\alpha\beta})^2-\rho^2,\nonumber \\ \tilde
k^{\alpha\beta}&=&(k_F)^{\alpha\mu\beta\nu}\hat p_\mu \hat p_\nu
\, , \quad \hat p^\mu=\frac{p^\mu}{|\vec p|}.
\end{eqnarray}
Then the dispersion relation is \cite{KosteleckyMewesPRD}
\begin{equation}
k^0_\pm=(1+\rho\pm\sigma)|\vec k|.
\end{equation}
The last term in this relation, which is proportional to $\sigma$,
is purely polarization--dependent. Astrophysics shows that such a
dependence, if it exists, is well below the levels relevant here
\cite{KosteleckyMewes}. We can thus assume $\sigma=0$.

To obtain the explicit time--dependence of the wavevector in our
experiment, we need to transform the quantities, which are
conventionally defined in a sun--centered celestial equatorial
reference frame  frame into the laboratory frame
\cite{KosteleckyMewesPRD}.

Explicit values of the Lorentz violating quantities depend on the
definition of coordinates and fields
\cite{KosteleckyMewesPRD,Mueller03,Mueller07,Mueller08}. The
freedom to define these can be used to set certain components to
zero. A particular definition, for example could be made by
requiring $\tilde\kappa_{e-}^{JK}=0$ (see appendix
\ref{EMappendix}). A choice like this would in general make these
terms reappear in other sectors, such as the gravitational and
fermionic sectors. In the following, we do not make any particular
assumptions on such definitions and retain all the quantities
$\bar s$ and $\tilde\kappa_{e-},\tilde\kappa_{o+}$ in full
generality.

\subsubsection{Combined signal}

Adding the contributions of the electromagnetic and the
gravitational sector, the time--dependence of the interferometer
phase can be expressed as a Fourier series for the time-dependence
\cite{BaileyKostelecky}
\begin{equation}\label{Fourierseries} \frac{\delta
\varphi}{\varphi_0}= \sum_m C_m\cos(\om_m t+\phi_m)+D_m\sin(\om_m
t+\phi_m).
\end{equation}
The coefficients $C_m, D_m$ for the six frequencies $m\in\{
\oo,2\oo,\oo\pm \Om,2\oo\pm \Om\}$ are functions of the Lorentz
violations, that are given by the components of $\bar s^{\mu\nu}$
and $(k_F)_{\kappa\lambda\mu\nu}$ and the frequencies of Earth's
orbit $\Oo=2\pi/(1$\,y) and rotation $\om_\oplus\simeq
2\pi/(23.93$\,h). For a vertical interferometer, we obtain the
signals for Lorentz violation in electromagnetism and gravity, see
Tab. \ref{VertInterf}. We denoted $i_4=1-3I_\oplus/(M_\oplus
R_\oplus^2)\approx -1/2$. The symbols
$\tilde\kappa_{e-},\tilde\kappa_{o+}$ denote linear combinations
of the elements of $(k_F)_{\kappa\lambda\mu\nu}$ that are defined
in appendix \ref{EMappendix}.

It turns out that for these components (but not, for example, for
$C_\Om, D_\Om$ or the ones in Tab. \ref{Hamplitudes}), the
substitutions
\begin{equation}
i_4\sigma^{JK}=\is^{JK}-\kemin^{JK},\quad
i_4\sigma^{TJ}=\is^{TJ}+\frac12 \epsilon_{JKL}\kopl^{KL}
\end{equation}
can be used to obtain the combined signal components from the
purely gravitational ones listed in Tab. IV of Ref.
\cite{BaileyKostelecky}.

\begin{table}
\caption{\label{VertInterf} Signal components for vertical atom
interferometers. $\chi$ is geographical colatitude, $42.3^\circ$
for Stanford.}
\begin{tabular}{ccc} \hline
Comp. & Amplitude & Phase \\ \hline \hline $C_{2\om}$ &
$\frac14\sin^2\chi[i_4(\bar s^{XX}-\bar
s^{YY})-(\kemin^{XX}-\kemin^{YY})]$ & $2\phi$ \\
$D_{2\om}$ & $\frac12 \sin^2\chi(\is^{XY}-\kemin^{XY})$ & $2\phi$ \\
$C_\om$ & $\frac12\sin2\chi(\is^{XZ}-\kemin^{XZ})$ & $\phi$ \\
$D_\om$ & $\frac12\sin2\chi(\is^{YZ}-\kemin^{YZ})$ & $\phi$ \\
$C_{2\om+\Om}$ & $-\frac14(\cos\eta-1)V_\oplus\sin^2\chi(\is^{TY}-\kopl^{XZ})$ & $2\phi$\\
$D_{2\om+\Om}$ &
$\frac14(\cos\eta-1)V_\oplus\sin^2\chi(\is^{TX}+\kopl^{YZ})$ &
$2\phi$
\\
$C_{2\om-\Om}$ & $-\frac14(\cos\eta+1)V_\oplus\sin^2\chi(\is^{TY}-\kopl^{XZ})$ & $2\phi$\\
$D_{2\om-\Om}$ &
$\frac14(\cos\eta+1)V_\oplus\sin^2\chi(\is^{TX}+\kopl^{YZ})$ &
$2\phi$
\\
$C_{\om+\Om}$ & $\frac14V_\oplus\sin\eta\sin^2\chi(\is^{TX}+\kopl^{YZ})$ & $\phi$\\
$D_{\om+\Om}$ &
$\frac14V_\oplus\sin^2\chi[(1-\cos\eta)(\is^{TZ}+\kopl^{XY})$ & $\phi$\\
$\pt{\om-\Om}$ & $ -\sin\eta(\is^{TY}-\kopl^{XZ})]$ &
\\
$C_{\om-\Om}$ & $\frac14V_\oplus\sin\eta\sin^2\chi(\is^{TX}+\kopl^{YZ})$ & $\phi$\\
$D_{\om-\Om}$ &
$\frac14V_\oplus\sin^2\chi[(1+\cos\eta)(\is^{TZ}+\kopl^{XY})$ & $\phi$ \\
$\pt{\om-\Om}$ & $+\sin\eta(\is^{TY}-\kopl^{XZ})]$ &
\\ \hline \hline
\end{tabular}
\end{table}

\subsection{Data analysis} \label{deconvolution}

\begin{figure*}
\centering \epsfig{file=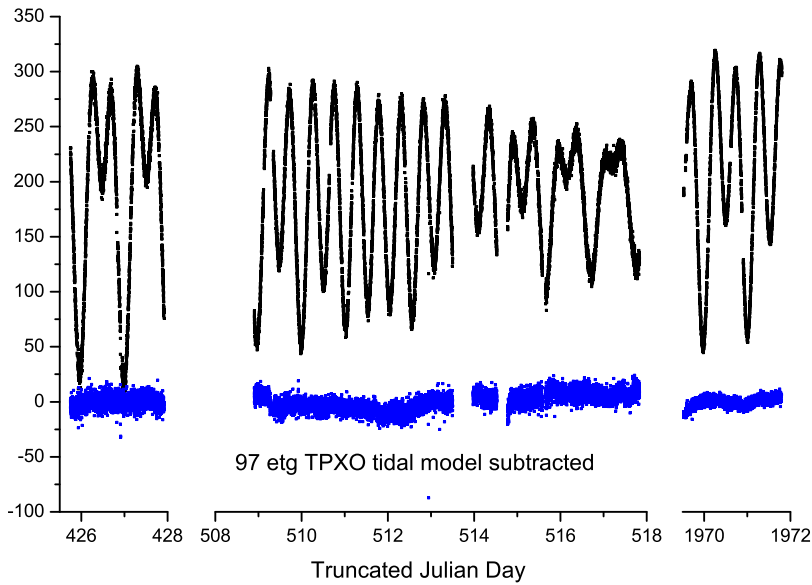, width=0.6\textwidth}
\caption{Data in $10^{-9}g$. Each point represents a 60-s scan of
one fringe (75-s after TJD1900). \label{alldata}}
\end{figure*}

The combined data spans about 1500\,d, but fragmented into three
relatively short segments: $\sim$60\,h of data taken with this
setup, as well as a $\sim 60$\,h and a $\sim$10\,d run reported
previously \cite{Peters}, see Fig. \ref{alldata}.

Major systematic effects in this experiment are tidal variations
of the local gravitational acceleration. Subtraction of a
Newtonian model of these influences \cite{Wenzel} based on the
relative positions of the Sun, the Moon, and the planets, yields
the residues shown at the bottom of Fig. \ref{alldata}. In
addition to the tidal model used in our Letter, we here use an
additional model of the local tides \cite{Egbert}. This will allow
us to significantly reduce the relatively large estimates on some
of the $\sigma$ coefficients reported in the Letter. As our old
analysis and analyses using lunar laser ranging \cite{Battat}, the
present analysis is based on an assumption of no accidental
cancelations of signals, i.e., that the model does not contain
the same influence of Lorentz violation as our measurement.

In order to extract the signal components of Tab.
\ref{VertInterf}, we calculate the discrete Fourier transform. As
is typical of a finite set of data, the Fourier components
overlap. This overlap can be quantified by a covariance matrix.
For a compact presentation, let
\begin{eqnarray}
\omega_{m}&=&(2\oo,\oo,2\oo+\Om,2\oo-\Om,\oo+\Om,\oo-\Om)\nonumber
\\
a_{2m+1}&=&\cos(\omega_m t+\phi_m), \quad
a_{2m}=\sin(\omega_mt+\phi_m)\,.
\end{eqnarray}
For a total of $K$ data, let $d(t_k), k\in \{1, \ldots K\}$ be the
datum taken at the time $t_k$. The sine and cosine Fourier
transforms are
\begin{equation}
\tilde d_m=\frac 2K\sum_k d(t_k) a_m(t_k).
\end{equation}
the $\tilde d_m$ that are obtained from the fragmented data are a
linear combination
\begin{eqnarray}
\tilde d_m= \sum_n \cov(a_m,a_n) \tilde D_n.
\end{eqnarray}
of the corresponding Fourier components $\tilde D_m$ for
hypothetical continuous data. The covariance matrix
\begin{equation}
\cov(a_m,a_n)=\frac 2K \sum_k a_m(t_k) a_n(t_k),
\end{equation}
where the sum is over all $t_k$. For data that span a large
time-scale, it approaches a unit matrix, i.e., the overlap of the
Fourier coefficients becomes negligible. The covariance matrix for
our data is shown in Tab. \ref{cor}.

\begin{table}
\centering \caption{\label{cor} Correlation matrix $\cov$.}
\begin{tabular}{cccccccccccc}\hline\hline
1& 0.& 0.& 0.& 0.3& -0.6& 0.3& 0.5& 0.& 0.& 0.1& 0.\\0.& 1& 0.&
0.& 0.6&
    0.3& -0.6& 0.3& 0.& 0.& 0.& 0.\\0.& 0.& 1& 0.& 0.& -0.1& 0.& 0.1& 0.6& -0.6&
    0.6& 0.6\\0.& 0.& 0.& 1& 0.& 0.& 0.& 0.& 0.6& 0.6& -0.6& 0.6\\0.3& 0.6& 0.&
    0.& 1& 0.& -0.6& 0.7& 0.& 0.& 0.& 0.\\-0.6& 0.3& -0.1& 0.& 0.& 1& -0.7& -0.6&
    0.& 0.& 0.& 0.\\0.3& -0.6& 0.& 0.& -0.6& -0.7& 1& 0.& 0.& 0.& 0.& 0.\\0.5& 0.3&
    0.1& 0.& 0.7& -0.6& 0.& 1& 0.& -0.1& 0.& 0.\\0.& 0.& 0.6& 0.6& 0.& 0.& 0.& 0.& 1&
     0.& 0.3& 0.6\\0.& 0.& -0.6& 0.6& 0.& 0.& 0.& -0.1& 0.& 1& -0.6& 0.3\\0.1&
    0.& 0.6& -0.6& 0.& 0.& 0.& 0.& 0.3& -0.6& 1& 0.\\0.& 0.& 0.6& 0.6& 0.& 0.& 0.& 0.&
    0.6& 0.3& 0.& 1 \\ \hline\hline
\end{tabular}
\end{table}

Multiplication of the covariance matrix (Tab. \ref{cor}) with the
signals listed in Tab. \ref{VertInterf} (written as a vector
$(C_{2\om}, D_{2\om}, C_\om,\ldots , D_{\om-\Om})$) gives the
linear combinations that the experiment is sensitive to, see Tab.
\ref{linearcoeff}. For obtaining bounds on Lorentz violation, we
perform a numerical Fourier analysis of the data for the 12
components $c_m, d_m$, see the table. The error is estimated by
performing a Fourier analysis at several frequencies above and
below the signal frequencies and computing the root of the mean
square.

\begin{table*}
\caption{\label{linearcoeff} Signal components as obtained by a
numerical Fourier transform of the data. They correspond to linear
combinations of the components of $\sigma$, with linear
coefficients as tabulated. The result of the Fourier transform is
listed as fraction of $g$.}
\begin{tabular}{ccccccccc}\hline
Comp. & $\sigma^{TX}$ & $\sigma^{TY}$ & $\sigma^{TZ}$ &
$\sigma^{XX}-\sigma^{YY}$ & $\sigma^{XY}$ & $\sigma^{XZ}$ & $\sigma^{YZ}$ & Meas. $/10^{-9}$ \\
\hline \hline $c_{2\om}$ & -0.091 & 0.040 &- 0.002 &
- 0.079 & 0.004 & - 0.009 & 0.004 & $-0.097\pm 0.24$ \\
$d_{2\om}$ & -0.039 & - 0.094 & - 0.007 & 0.001 & -
 0.241 & - 0.007 & 0.001 & $0.009\pm0.24$ \\
 $c_\om $& -0.066 & 0.001 & - 0.142 & - 0.003 & -
  0.007 & - 0.240 &  0.001 & $0.69\pm0.27$ \\
$d_\om$ & 0.003 & - 0.063 & - 0.136 &  0.001 &
  0.001 &  0.001 & - 0.243 & $-0.034\pm 0.27$ \\
$c_{2\om+\Om}$ &-0.105 & -
  0.094 & - 0.005 & - 0.020 & - 0.141 & -
  0.007 & - 0.004 & $0.19\pm0.24$ \\ $d_{2\om+\Om}$ & 0.098 & -
  0.107 &  0.011 &
0.044 & -
  0.068 &  0.016 & - 0.003 & $-0.22\pm 0.24$ \\ $c_{2\om-\Om}$ &-0.005 &
  0.158 & - 0.001& - 0.021 & 0.144 & -
  0.002 &  0.006 & $-0.15\pm 0.24$ \\ $d_{2\om-\Om}$ &-0.157 & - 0.000 & -
  0.004 & - 0.043 & - 0.064 & - 0.014 &
  0.006 & $0.40\pm0.24$ \\ $c_{\om+\Om}$ &-0.065 & - 0.034 & - 0.131 & -
  0.003 & - 0.002 & - 0.140 & - 0.143 & $0.20\pm0.27$ \\
  $d_{\om+\Om}$ &
 0.036 & - 0.064 & - 0.053 &  0.003&
  0.004 &  0.139 & - 0.151 & $-0.73\pm0.27$ \\ $c_{\om-\Om}$ &-0.067 &
  0.029 & - 0.003 & - 0.005 & -
  0.001 & - 0.143 &  0.140 & $0.28\pm 0.27$ \\ $c_{\om-\Om}$ &-0.029 & -
  0.061 & - 0.231 & - 0.001 & - 0.007 & -
  0.142 & - 0.148 & $0.25\pm 0.27$ \\ \hline\hline
\end{tabular}
\end{table*}

The limits listed in Tab. \ref{linearcoeff} are on linear
combinations of parameters. The 12 results are sufficient to
determine all 7 parameters. In order to obtain independent
estimates for the parameters while making optimum use of the
experimental data, we proceed as follows. Each datum corresponds
to a probability distribution
$p_n(\sigma^{TX},\ldots,\sigma^{YZ})$, where $n=1,\ldots 12$, that
we assume to be Gaussian with the center and standard deviation as
tabulated. Multiplying all 12 distributions results in one overall
probability distribution
$P(\sigma^{TX},\ldots,\sigma^{YZ})=A\prod_{n=1}^{12}p_n$, where
$A$ is a normalization factor. In order, for example, to obtain an
estimate on $\sigma^{TX}$ that is independent of the other
coefficients, we integrate over the other variables,
$P(\sigma^{TX})=A'\int_{-\infty}^\infty
d\sigma^{TY}\ldots\int_{-\infty}^\infty d\sigma^{YZ}
P(\sigma^{TX},\ldots,\sigma^{YZ})$ (where all variables but
$\sigma^{TX}$ are integrated over and $A'$ is another
normalization factor). The result is a probability distribution
for $\sigma^{TX}$ that has one maximum and standard deviation,
which are our estimate and error bar for this coefficient.
Specifically, since we assumed the $p_n$ to be Gaussian,
$P(\sigma^{TX})$ will be a Gaussian and the most probable estimate
as well as one $\sigma$ error are read off.

Tab. \ref{finalbounds} shows the results thus obtained. Compared
to the ones that we published in our Letter \cite{LVGrav} (derived
from the same data), the limits are now more uniform: whereas the
best limits on the three $\sigma^{TJ}$ and four (combinations of)
$\sigma^{JK}$ are of the same order, the worst ones from the new
analysis are about four times better than in the Letter. This is a
result of our use of a more sophisticated tidal model as well as
an optimum method to derive individual limits from the limits on
linear combinations.

\begin{table}
\centering\caption{\label{finalbounds}}
\begin{tabular}{cc}\hline
Coeff. & \\ \hline\hline
$\sigma^{TX}$ & $(-3.1\pm 5.1)\times 10^{-5}$ \\
$\sigma^{TY}$ & $(0.1\pm 5.4)\times 10^{-5}$ \\
$\sigma^{TZ}$ & $(1.4\pm 6.6) \times 10^{-5}$ \\
$\sigma^{XX}-\sigma^{YY}$ & $(4.4\pm11)\times 10^{-9}$  \\
$\sigma^{XY}$ & $(0.2\pm 3.9)\times 10^{-9}$ \\
$\sigma^{XZ}$ & $(-2.6\pm 4.4)\times 10^{-9}$ \\
$\sigma^{YZ}$ & $(-0.3\pm 4.5)\times 10^{-9}$ \\
\hline\hline
\end{tabular}
\end{table}

\subsection{Combination with lunar laser ranging (LLR)}

Our experiment thus provides the only measurement of the
components of $\sigma$ so far; these are combinations of
gravitational and electromagnetic Lorentz violation. To compare
our data with previous experiments, which have not been analyzed
for the electromagnetic influence (although such an influence
exist), we here adopt the assumption that there is no Lorentz
violation in electromagnetism, i.e., $\tilde
\kappa_{e-}^{JK}=\tilde \kappa_{o+}^{JK}=0$. Our limits (Tab.
\ref{finalbounds}) then correspond directly to bounds on $\bar s$.

Lunar laser ranging \cite{Battat} bounds two linear combinations
of $\bar s^{JK}$ \cite{Battat}
\begin{eqnarray}
\bar s^{11} - \bar s^{22} &\equiv& 0.08(\bar s^{XX} + \bar s^{YY}
- 2\bar s^{ZZ}) \nonumber \\ && -0.31(\bar s^{XX} -\bar s^{YY})  -
1.7\bar s^{XY} \nonumber \\ && +0.60\bar s^{XZ} + 0.42\bar
s^{YZ}\nonumber \\ & =&(1.3\pm0.9)\times 10^{-10},\nonumber\\ \bar
s^{12}& \equiv& 0.43(\bar s^{XX} - \bar s^{YY}) - 0.31\bar s^{XY}
\nonumber
\\ &&-0.23\bar s^{XZ} - 0.33 \bar s^{YZ}\nonumber \\ &=&(6.9\pm 4.5)\times
10^{-11}.
\end{eqnarray}
They can be combined with our results from Tab. \ref{linearcoeff}
(using the method described above) to obtain independent limits on
one more degree of freedom of $\bar s^{JK}$. Battat {\em et al.}
also report four limits on the three $\bar s^{TJ}$, \cite{Battat}
\begin{eqnarray}
\bar s^{01} &=& -0.60\bar s^{TX} + 0.82\bar s^{TY} \nonumber \\ &=&(-0.8\pm 1.1)\times 10^{-6} ,\nonumber \\
\bar s^{02} &=& -0.53\bar s^{TY} - 0.75\bar s^{TX} + 0.40\bar
s^{TZ}\nonumber \\ &=&(-5.2\pm 4.8)\times 10^{-7}, \nonumber  \\
\bar s_{\Om_\oplus c} &=&
-3.1\bar s^{TY}- 1.1\bar s^{TZ} + 0.094\bar s^{TX}\nonumber \\ &=&(0.2\pm 3.9)\times 10^{-7},\nonumber \\
\bar s_{\Om_\oplus s} &=& -3.4\bar s^{TX} + 0.037\bar s^{TY} +
0.15\bar s^{TZ}\nonumber \\ &=&(-1.3\pm 4.1)\times 10^{-7}.
\end{eqnarray}
They can be combined with ours to increase the resolution of the
limits.

Tab. \ref{combinedbounds} lists the results thus obtained. They
represent the most complete bounds on Lorentz violation in
gravity, providing individual limits on the $\bar s$ as well as
more components of $\bar s$ and higher resolution than either
experiment. The only degrees of freedom of $\bar s^{JK}$ that are
not bounded are $\bar s^{TT}$ and the trace, which do not lead to
signals to first order in the Earth's orbital velocity. 

\begin{table}
\centering\caption{\label{combinedbounds} Bounds resulting from
combining our data with the ones from lunar laser ranging as
reported by Battat {\em et al.} \cite{Battat}, assuming vanishing
Lorentz violation in electrodynamics.}
\begin{tabular}{cc}\hline
Coeff. & \\ \hline\hline
$\bar s^{TX}$ & $(0.5\pm 6.2)\times 10^{-7}$ \\
$\bar s^{TY}$ & $(0.1\pm 1.3)\times 10^{-6}$ \\
$\bar s^{TZ}$ & $(-0.4\pm 3.8) \times 10^{-6}$ \\
$\bar s^{XX}-\bar s^{YY}$ & $(-1.2 \pm1.6)\times 10^{-9}$  \\
$\bar s^{XX}+\bar s^{YY}-2\bar s^{ZZ}$ & $(1.8\pm 38)\times
10^{-9}$ \\
$\bar s^{XY}$ & $(-0.6\pm 1.5)\times 10^{-9}$ \\
$\bar s^{XZ}$ & $(-2.7\pm 1.4)\times 10^{-9}$ \\
$\bar s^{YZ}$ & $(0.6\pm 1.4)\times 10^{-9}$ \\
\hline\hline
\end{tabular}
\end{table}

\section{Signal for horizontal interferometers}
\label{Hinterf}

In this section, we consider tests of gravity with horizontal atom
interferometers, including guided atom devices. Testing LLI in
gravity is a task that makes good use of the features of such
interferometers, in particular long coherence times and hence high
resolution. Moreover, since the signal for violations is a
time--dependent modulation, the stability of the interferometer on
time scales much larger than the modulation frequencies are not a
primary concern for such tests.

For simplicity, we shall again assume a vanishing of Lorentz
violation in electrodynamics throughout this section.

A test of the LLI of gravity can be performed by measuring a
Lorentz-violating horizontal acceleration. These accelerations are
given by \cite{BaileyKostelecky}
\begin{eqnarray}
a^x&=&-gi_3\bar s^{xz}-\oo^2R_\oplus\sin\chi\cos\chi\nonumber \\
&&+gi_3\bar s^{Tz}+gi_3\bar s^{Tx}V_\oplus^z,\nonumber\\
a^y&=&-gi_3\bar s^{yz}+gi_3\bar s^{Tz}+gi_3\bar
s^{Ty}V_\oplus^z,\label{Haccel}
\end{eqnarray}
where $i_3=1-I_\oplus/(M_\oplus R_\oplus^2)\approx 1/2$. For the
purpose of this section, we can take $a^z=g$ as well as $a\equiv
|a|\approx g$ to be constant. Such tests can, for example, be
based on a torsion pendulum, which is suspended off its center of
mass. Nevertheless, they might reach superior sensitivity compared
to vertical gravimeters such as the atom interferometer discussed
previously.

Compared to a conventional torsion pendulum, such experiments
involve special challenges associated with maintaining the
pendulum within the horizontal plane. However, measurement of the
horizontal accelerations that are given by Eqs. (\ref{Haccel})
with atom interferometry is possible using conventional horizontal
interferometers. Moreover, interferometers in the horizontal plane
can be built well using atom-chip or atomic waveguide techniques.
In contrast to atomic fountains, they allow long pulse separation
times $T_p$ in a compact setup. It is therefore interesting to
study the signals for Lorentz violation in post-Newtonian gravity
and electromagnetism for such an interferometer.

We assume a horizontal Mach-Zehnder interferometer with the laser
beams pointing into a direction of
\begin{equation}\label{AIorient}
\hat x=(\cos \theta,\sin\theta,0)
\end{equation}
in the laboratory frame. As before, the phase shift is given by
$k_{\rm eff} T_p^2(\hat x \cdot \vec g)$, where the local
gravitational acceleration $\vec g$ has vertical as well as
horizontal components.

The calculation of the induced time-dependence of the
interferometer phase proceeds via the transformations between the
laboratory frame and the sun--centered standard frame. The fastest
way to do this is probably by analogy to the case of torsion
balances that has been considered in \cite{BaileyKostelecky}; see
appendix \ref{Horappendix}. After all, both measure the
accelerations given by Eqs. (\ref{Haccel}). As a result, we can
express the contribution of Lorentz violation in gravity to the
phase as
\begin{eqnarray}
\varphi=k_{\rm
eff}i_3gT_p^2\sum_n\left([E_n\sin\alpha_n-F_n\cos\alpha_n]\sin\om_nT\right.\nonumber
\\ \left.
-[E_n\cos\alpha_n+F_n\sin\alpha_n]\cos\om_nT\right).\quad
\end{eqnarray}
The amplitudes and phases in this expression are given in Tab.
\ref{Hamplitudes}. It is evident that horizontal interferometers
provide access to four independent linear combinations of $\bar
s^{JK}$ (the same ones as vertical interferometers) and sufficient
data to determine all the $\bar s^{TJ}$.

\begin{table*}
\centering \caption{\label{Hamplitudes} Amplitude and phase for
the gravitational part of the signal in horizontal
interferometers}
\begin{tabular}{ccc} \hline
& Amplitude & $\alpha_n$ \\
\hline \hline $E_{2 \om} $ & $ -\overline{s}^{XY} \sin\theta \sin
\chi - \frac 14 (\overline{s}^{XX}-\overline{s}^{YY})
\cos\theta \sin 2 \chi $ & $2\phi$ \\
$F_{2 \om} $ & $ \frac 12 (\overline{s}^{XX}-\overline{s}^{YY})
\sin\theta \sin \chi
- \frac 12 \overline{s}^{XY} \cos\theta \sin 2\chi $ & $2\phi$ \\
$E_{\om} $ & $ -\overline{s}^{YZ} \sin\theta \cos \chi
- \overline{s}^{XZ} \cos\theta \cos 2\chi $ & $\phi$ \\
$F_{\om} $ & $ \overline{s}^{XZ} \sin\theta \cos \chi
- \overline{s}^{ZY} \cos\theta \cos 2\chi $ & $\phi$ \\
$E_{2\om + \Om} $ & $  \frac 12 V_{\oplus} \overline{s}^{TX}
\sin\theta (1-\cos \eta) \sin \chi - \frac 14 V_{\oplus}
\overline{s}^{TY} \cos\theta (1- \cos \eta)  \sin 2\chi$
& $2\phi$ \\
$F_{2\om + \Om} $ & $ \frac 12 V_{\oplus} \overline{s}^{TY}
\sin\theta (1- \cos \eta) \sin \chi + \frac 14 V_{\oplus}
\overline{s}^{TX} \cos\theta (1-\cos \eta) \sin 2\chi$
& $2\phi$ \\
$E_{2\om - \Om} $ & $ -\frac 12 V_{\oplus} \overline{s}^{TX}
\sin\theta (1 + \cos \eta) \sin \chi + \frac 14 V_{\oplus}
\overline{s}^{TY} \cos\theta (1+ \cos \eta) \sin 2\chi $
& $2\phi$ \\
$F_{2\om - \Om} $ & $ -\frac 12 V_{\oplus} \overline{s}^{TY}
\sin\theta ( 1 + \cos \eta) \sin \chi -\frac 14 V_{\oplus}
\overline{s}^{TX} \cos\theta (1+ \cos \eta) \sin 2\chi $
& $2\phi$ \\
$E_{\om + \Om} $ & $ -\frac 12 V_{\oplus} \overline{s}^{TY}
\sin\theta \sin \eta \cos \chi + V_{\oplus} \overline{s}^{TX}
\cos\theta \sin \eta (\frac 12 - \cos^2 \chi)$
& $\phi$ \\
$\pt{E_{\om + \Om}} $ & $ -\frac 12 V_{\oplus} \overline{s}^{TZ}
\sin\theta (\cos \eta -1) \cos \chi $
& \\
$F_{\om + \Om} $ & $  \frac 12 V_{\oplus} \overline{s}^{TX}
\sin\theta \sin \eta \cos \chi - V_{\oplus} \overline{s}^{TZ}
\cos\theta (1 - \cos \eta)(\frac 12-\cos^2 \chi)$
& $\phi$ \\
$\pt{F_{\om + \Om}} $ & $ + V_{\oplus} \overline{s}^{TY}
\cos\theta \sin \eta (\frac 12-\cos^2 \chi) $
& \\
$E_{\om - \Om} $ & $ -\frac 12 V_{\oplus} \overline{s}^{TZ}
\sin\theta (1+\cos \eta ) \cos \chi +\frac 12 V_{\oplus}
\overline{s}^{TY} \sin\theta \sin \eta \cos \chi$
& $\phi$ \\
$\pt{E_{\om - \Om}} $ & $ + V_{\oplus} \overline{s}^{TX}
\cos\theta \sin \eta (\frac 12-\cos^2 \chi) $
& \\
$F_{\om - \Om} $ & $ \frac 12 V_{\oplus} \overline{s}^{TX}
\sin\theta \sin \eta \cos \chi + V_{\oplus} \overline{s}^{TY}
\cos\theta \sin \eta (\frac 12-\cos^2 \chi)$
& $\phi$ \\
$\pt{F_{\om - \Om}} $ & $ + V_{\oplus} \overline{s}^{TZ}
\cos\theta (1+ \cos \eta)(\frac 12-\cos^2 \chi)$
& \\
$E_\Om $ & $ - \frac 12 V_{\oplus} \overline{s}^{TY} \cos\theta
\cos \eta \sin 2\chi
+ V_{\oplus} \overline{s}^{TZ} \cos\theta \sin \eta \sin 2\chi $ & $0$ \\
$F_\Om $ & $
\frac 12 V_{\oplus} \overline{s}^{TX} \cos\theta \sin 2 \chi $ & $0$ \\
\hline\hline
\end{tabular}
\end{table*}

\section{Summary and Outlook}

In this paper, we have presented atom-interferometry tests of the
local Lorentz invariance of post-Newtonian gravity and
electrodynamics. As a comprehensive and quantitative model for
violations, we use the standard model extension
\cite{ColladayKostelecky,Kostelecky}. The relevant violations of
LLI in gravity are encoded by a tensor $\bar s$
\cite{BaileyKostelecky}; those for violations in electromagnetism
are expressed by the matrices $\tilde\kappa_{e-}$ and
$\tilde\kappa_{o+}$ \cite{KosteleckyMewes}. The experimental
signal for Lorentz violation is a time-dependence of the local
gravitational acceleration as the Earth orbits in the solar
system. We discuss an experiment that has been performed by us
\cite{LVGrav} as well as possible future experiments with
horizontal interferometer geometries.

Our experiment is a vertical Mach-Zehnder atom interferometer. It
uses a bright source of cesium atoms in a 1-m high atomic fountain
and a pulse separation time of $400\,$ms. A resolution of up to
$8\times 10^{-9}\,g/\sqrt{\rm Hz}$ is reached, the highest of any
cold-atom gravimeter so far.

For this experiment, the signal for Lorentz violation is given by
a particular combination of coefficients entering the
gravitational and electromagnetic sectors: $\sigma^{JK}=\bar
s^{JK}-\kemin^{JK}/i_4$ and $ \sigma^{TJ}=\bar
s^{TJ}+\epsilon_{JKL}\kopl^{KL}/(2i_4)$. Here, $i_4\simeq -1/2$ is
given by the Earth's spherical moment of inertia.

For the analysis, we use about 2.5\,d of data taken with the setup
just described as well as about 12.5\,d of data taken with a
previous setup \cite{Peters}. The data have been taken over a
total time interval of about 1500\,d. A major systematic effect
are the tides, which we account for by subtracting a Newtonian
model that is based on the relative positions of the Earth, the
Moon, the Sun, and planets. By Fourier analyzing the residuals at
different combinations of the frequencies of Earth's rotation and
orbit, we obtain bounds on twelve linear combinations of seven
components of $\sigma^{JK}$ at the $10^{-9}$ level.

Our limits are the strongest bounds on several combinations of
coefficients for Lorentz violation, even in view of previous work
from geophysical gravity observations \cite{Nordtvedt76} and lunar
laser ranging \cite{Battat}. This makes our experiment one of the
few competitive laboratory tests of gravity. Moreover, our
experiment is presently the only test of Lorentz invariance of
gravity that explicitly takes into account the possibility of
Lorentz violation in the non-gravitational standard model. This is
important when interpreting the results in the context of the
search for quantum-gravity signals.

The physics behind atom interferometry and lunar laser ranging is
similar, monitoring the trajectory of a proof mass within Earth's
gravitational field by laser beams. However, they differ in the
orbit (if one can think of the atoms' trajectory as an orbit) and
quantum-mechanical nature of the mass. Indeed, the equivalence
principle for quantum objects has been discussed \cite{Herdegen}.
Also, ours is a laboratory experiment, which typically offers
superior control over the experimental systematics. Moreover, it
is so far the only experiment where the simultaneous influence of
the non-gravitational and gravitational effects are understood
quantitatively and which accordingly states combined bounds.

Assuming that Lorentz invariance in electrodynamics vanishes, the
data from LLR and our experiment can be combined to yield improved
and more detailed bounds.

We also studied the signals for Lorentz violation (in gravity and
electromagnetism) for horizontal interferometers. They are
attractive for this type of measurement, as they offer increased
coherence time in a compact setup.

Taking several months worth of data would help to eliminate the
dominant lunar tides, the major systematic effect. With a dataset
that spans a year, even the leading solar tides could be
suppressed.

Future work may lead to alternative bounds from torsion balances, gravimetry data that is routinely taken in geophysical research,
or an analysis of data from the gravity probe-B satellite
\cite{James}. Note, however, that our results are still not
limited by the any fundamental influence such as quantum
projection noise. With typically $10^8$ atoms per launch, a
quantum projection limited gravimeter could reach the $10^{-12}g$
level per launch and $10^{-14}g$ per day, if other noise sources
(notably phase noise and vibrations) can be controlled. Also, the
performance of atom interferometers can be increased further by
utilizing beam splitters that transfer the momentum of many
photons, thus leading to sensitivity increases by factors of 12
(demonstrated) to 100 times (anticipated)
\cite{BraggInterferometry,SCI,BBB}. This promises improved tests
of gravity based on atom interferometry, deepening our
understanding of the fundamental principles of Nature.

\acknowledgements We thank A. Peters and A. Senger for valuable
help and important discussions.

\appendix

\section{Electromagnetic sector}\label{EMappendix}
The Maxwell equations in vacuum that are derived from the
Lagrangian, Eq. (\ref{SMEphotonlagrangian}), are
\begin{eqnarray}\label{inhomMaxwell}
\partial_\alpha F^\alpha_\mu + (k_F)_{\mu \alpha \beta \gamma} \partial^\alpha F^{\beta \gamma}&
=&
0,\nonumber \\
\partial_\mu \tilde F^{\mu \nu} &=& 0,
\end{eqnarray}
where
\begin{equation}
\tilde F^{\mu \nu} = \frac12
\varepsilon^{\mu\nu\alpha\beta}F_{\alpha\beta}.
\end{equation}
They can be written in analogy to the Maxwell equations in
anisotropic media \cite{KosteleckyMewesPRD}: With the $3 \times 3$
matrices
\begin{equation}\label{kappadef}
\begin{array}{ll} (\kappa_{DE})^{jk} = -2 (k_F)^{0j0k}, & (\kappa_{HB})^{jk} = \frac 12
\epsilon^{jpq} \epsilon^{krs} (k_F)^{pqrs}, \\ & \\
(\kappa_{DB})^{jk} = (k_F)^{0jpq} \epsilon^{kpq}, &
(\kappa_{HE})^{kj} = - (\kappa_{DB})^{jk} \nonumber
\end{array}
\end{equation}
(latin indices take the values $1,2,3$) one can define $\vec D$
and $\vec H$ fields
\begin{equation}
\left( \begin{array}{c} \vec D \\ \vec H \end{array} \right) =
\left( \begin{array}{cc} \openone + \kappa_{DE} & \kappa_{DB} \\
\kappa_{HE} & \openone + \kappa_{HB} \end{array} \right) \left(
\begin{array}{c} \vec E \\ \vec B \end{array} \right),
\end{equation}
where $\openone$ represents the $3 \times 3$ unit matrix. The
Maxwell equations can now be expressed as
\cite{KosteleckyMewesPRD}
\begin{equation}
\begin{array}{ll} \vec \nabla \times \vec H - \partial_0
\vec D = 0, & \vec \nabla \cdot \vec B = 0,
\\ & \\ \vec \nabla \times \vec E +
\partial_0 \vec B = 0, & \vec \nabla \cdot \vec D = 0. \nonumber
\end{array}
\end{equation}
Lorentz violation in electrodynamics is thus analogous to
electrodynamics in anisotropic media. For later use, we define the
linear combinations
\begin{eqnarray}\label{tildekappas}
(\tilde \kappa_{e+})^{jk} & = & \frac 12
(\kappa_{DE}+\kappa_{HB})^{jk}, \nonumber
\\ (\tilde \kappa_{o+})^{jk}& = & \frac 12
(\kappa_{DB}+ \kappa_{HE})^{jk}, \nonumber \\ (\tilde
\kappa_{e-})^{jk} & = & \frac 12 (\kappa_{DE}-\kappa_{HB})^{jk} -
\frac 13 \delta^{jk} (\kappa_{DE})^{ll},\nonumber \\ (\tilde
\kappa_{o-})^{jk} & = & \frac 12 (\kappa_{DB}-\kappa_{HE})^{jk},
\nonumber \\ \tilde \kappa_{tr}&=&\frac13(\kappa_{DE})^{ll}.
\end{eqnarray}
Of these, the ten degrees of freedom of $\tilde\kappa_{o-}$ and
$\tilde \kappa_{e+}$ encode birefringence; they are bounded to
below $10^{-37}$ by observations of gamma-ray bursts
\cite{KosteleckyMewes,KosteleckyMewesPRD}. The residual nine cause
a dependence of the velocity of light on the direction of
propagation or, vice versa, such a dependence of the wavenumber of
light having a constant frequency. They are therefore relevant in
interferometry experiments.

\section{Analogy of horizontal interferometers and torsion
pendulum experiments} \label{Horappendix}

The free oscillations of a torsion pendulum are governed by the
differential equation $\ddot \vartheta+(\kappa/I)\vartheta=\tau$,
where $I$ is the moment of inertia, and $\tau$ represents the
driving force. It can be expressed by an effective potential $V$
as $\tau=-dV/d\vartheta$. In the limit of $\kappa\rightarrow
\infty$, we simply have
\begin{equation}
\theta(T)=-\frac1\kappa\frac{dV}{d\vartheta}.
\end{equation}
Kostelecky and Bailey \cite{BaileyKostelecky} give the
oscillations $\vartheta(T)$ by their Eq. (134). For small
excitations $\vartheta$,
\begin{equation}
\frac{dV}{d\vartheta}=-mr_0(a^xS_N-a^yC_N+\frac12a\om_\oplus^2\sin^2\chi
S_{N2}).
\end{equation}
Equating this with Bailey and Kostelecky's Eq. (134) in the limit
of $\om_0\rightarrow \infty$, we obtain
\begin{eqnarray}
a^xS_N-a^yC_N+\frac12a\om_\oplus^2\sin^2\chi S_{N2}\nonumber \\
=
i_3g\sum_n\left([E_n\sin\alpha_n-F_n\cos\alpha_n]\sin(\om_nT)\right.\nonumber
\\ \left. -[E_n\cos\alpha_n+F_n\sin\alpha_n]\cos(\om_nT)\right)
\end{eqnarray}
where the amplitudes $E_n, F_n$ and the phases $\alpha_n$ are
tabulated in Tab. V of \cite{BaileyKostelecky}. They represent the
time-dependence of the Lorentz-violating horizontal accelerations
as functions of the Earth's orbital parameters and the
$C_N,S_N,S_{2N}$. Specializing to the case of one mass,
$C_N=\cos\theta, S_N=\sin\theta,$ and $S_{2N}=\sin2\theta$. By
comparison of coefficients, we obtain explicit expressions for
$a_x, a_y,$ and $a_z$ in the laboratory frame, that we can insert
into the equation $\varphi=\vec k_{\rm eff} \vec g T_p^2$ for the
interferometer's phase. Note that with our specification of the
atom interferometer's orientation $\hat x$, Eq. (\ref{AIorient}),
$\theta =\vartheta+\pi/2$, which has been taken into account in
Tab. \ref{Hamplitudes}.

\end{document}